\documentclass[11pt,american,english]{article}
\usepackage[latin9]{inputenc}
\usepackage{float}
\usepackage{algorithm2e}
\usepackage{amsmath}
\usepackage{amssymb}
\usepackage{graphicx}

\makeatletter
\pdfoutput=1
\usepackage{latexsym}
\usepackage{graphicx}

\usepackage{newfloat}

\usepackage{url}
\usepackage{wrapfig}
\usepackage[bookmarks]{hyperref}
\usepackage{nicefrac}

\hypersetup{
    colorlinks=true,
    linkcolor=blue,
    urlcolor=blue,
    citecolor=blue,
}
\renewcommand*{\@fnsymbol}[1]{\ensuremath{\ifcase#1\or *\or \mathsection\or \mathparagraph\or \dagger\or \ddagger\or
      \|\or **\or \dagger\dagger
    \or \ddagger\ddagger \else\@ctrerr\fi}}

\DeclareFloatingEnvironment[placement={!ht},name=List]{mylist}

\makeatother

\usepackage{babel}

\begin{document}
\title{Mechanisms of intermediary platforms}
\author{Tobias K\"olbel \thanks{Project Economy of Things, Robert Bosch Manufacturing Solutions GmbH, Leitzstrasse 47, 70469 Stuttgart, Germany}\and Daniel Kunz \thanks{Project Economy of Things, Robert Bosch GmbH, Corporate Research Campus, 71272 Renningen, Germany;
The project underlying this report was partially funded by the German Federal Ministry of Education and Research (BMBF) iBlockchain project under grant number 16KIS0904. The responsibility for the content of this publication lies with the author.}}
\date{\the\year}

\maketitle
\begin{abstract}
    In  the  current  digital  age  of  the  Internet,  with  ever-growing  networks  and  data-driven  business models, digital platforms and especially marketplaces are becoming increasingly important. These platforms focus primarily on digital businesses by offering services that bring together consumers and producers. Due to added value created for consumers, the profit-driven operators of these platforms (\textit{Matchmakers}) are extremely successful and have come to dominate their respective markets.

    The aim of this article is to understand how Matchmakers and coordination networks gain market dominance. The following sections will take a closer look at network and coordination effects as well as intermediary platform mechanisms and entailing disadvantages for users. Considering strategic and business challenges, we suggest a possible solution and strategy to avoid dependencies on individual players in the digital economy. We present a cooperative approach towards a fair and open \textit{Economy of Things} (EoT) based on decentralized technologies. 
\end{abstract}

\section{Introduction}

"Uber, the world's largest taxi company owns no vehicles, Facebook, the world's most popular media owner creates no content, Alibaba, the most valuable retailer has no inventory and Airbnb, the world's largest accommodation provider owns no real estate. Something interesting is happening."

\hspace*{\fill}[Tom Goodwin 2015]

In the current digital age of the Internet, with ever-growing networks and data-driven business models, digital platforms and especially marketplaces are becoming increasingly important. 
These platforms focus primarily on digital businesses by offering services that bring together consumers and producers. Due to added value created for consumers, the profit-driven operators of these platforms are extremely successful, have come to dominate their respective markets and tech companies such as \textit{Amazon} are among the most valuable companies in the world \cite{statista_biggest_companies}.

For a better understanding, we present the most popular platforms in order to set boundaries for a closer look: \textit{Uber} brings together drivers and passengers by matching these two based on the route given by a possible passenger. \textit{Airbnb} creates matches between apartment owners and travelers for a given time range and location. \textit{Amazon Marketplace} matches for generic consumer goods between sellers and buyers. \textit{Alibaba} and \textit{eBay} offer systems based on auctions for any kind of good, serving both sellers and buyers.

Every of the described platforms offers a digital service that satisfies a specific customer's need. As a \textit{Multi-Sided Platform} (MSP), they act as intermediaries in order to bring together multiple types of groups. A distinction can be made between producers and consumers, sellers and buyers or more generic, supply and demand for each type of service. All these MSPs -- also called \textbf{\textit{Matchmakers}} by \cite{evans_2016} -- have the following aspects in common:

\begin{itemize}
    \item MSPs are operated in markets with significant market frictions, economists tend to call them transaction costs \cite{evans_2016,rochet_platform_2003,reillier_platform_2017,hagiu_marketplace_2015} that prevent market participants from interacting easily and directly with each other and successful MSPs aim to reduce these frictions and connect willing buyers and sellers.
    \item They are transaction-based systems which balance the interdependent demands of multiple groups of participants and charge transaction fees for matchmaking between supply and demand; typically invoiced via the seller or producer side \cite{rochet_platform_2003,armstrong_competition_2006}.
    \item Matchmakers are in one hand, both organizationally (company) and operationally (platform curation).
    \item Their success is based on a growing marginal utility: the greater the offer, the more attractive the platform; the more attractive the platform, the more participants in the network. More participants in the network lead to a higher value or benefit of the platform for each participant. 
    Example: The more rides are offered by \textit{Uber}, the more attractive the platform becomes for riders. More travelers in turn attract more drivers. As a result of these causalities, the MSP is growing.
    \item MSPs provide information throughout the entire business process, starting with onboarding, search, continuing with the coordination  or matchmaking and finally the settlement. In detail, the entire market behavior is transparent for one side, e.g. what has been bought at what price and from whom.
\end{itemize}

From the past years, it can be seen that these platforms are very attractive for users and that they gain more and more power as their number of participants increases. This is due to self-reinforcing effects (\textit{network effects}), which lead to dependencies for  users (\textit{lock-in effects}), as it is more difficult for participants to switch to alternative platforms.

Once a platform reaches a critical mass of users, it becomes increasingly dominant and tends to establish a \textit{de facto} monopoly for its target domain over time. Matchmakers are able to dominate and dictate their market segment. Examples from the B2C sector illustrate these trends: \textit{Amazon} is expected to represent \nicefrac{1}{2} of the US e-commerce market by 2020 \cite{statista_us_amazon_market_share} and earns money by dealing with other partners -- albeit under the conditions dictated by \textit{Amazon}. \textit{Alibaba} controls 75\% of all chinese e-commerce transactions and 94\% of all online search queries are carried out via \textit{Google} \cite{bmwi_b2b_2019,vanalstyne_pipelines_2016}. Since their offer scales quickly and easily, growth rates can be assumed  exponential.

The aim of this article is to understand how matchmakers and coordination networks gain market dominance. The following sections will take a closer look at network and coordination effects as well as intermediary platform mechanisms and resulting disadvantages for users. Ultimately, we assume these are responsible for the rise of digital platform capitalism with a tendency towards \textit{de facto} monopolistic structures. Considering strategic and business challenges, we suggest a possible solution and strategy to avoid dependencies on individual players in the digital economy.

\section{\label{sec:network_effects}Network Effects on Platforms}

Users of various information technologies benefit from using a common format, system or network \cite{shapiro_information_1999}. All three have no significant value \textit{per se} -- their value is determined by the quality and quantity of the interactions they enable. 
If a product's value from a user's perspective depends proportionally on how many other users use the same product, economists say that this product has network effects (also called \textit{network externalities}) \cite{rochet_platform_2003,reillier_platform_2017,shapiro_information_1999,eisenmann_strategies_2006,hagiu_platforms_2015}.
Jeffrey Rohlfs (1974) first acknowledged this phenomenon in his pioneering paper\footnote{For a non-technical discussion, see \cite{schmalensee_introduction_2011}.}, which dealt with the beginnings of landline telephone services after the telephone was introduced \cite{rohlfs_comunication_1974}. Regarding his findings, a telephone was useless if nobody else had one. In fact, the more people a user could reach, the more valuable it would became.

\subsection{Classification of Network Effects}

If there are competing information technologies in the same market segment, they compete against each other to attract more and more users, and thus, create network effects. Their aim is to increase the value of its network based on the number of users. This is due to the self-reinforcing effect of network effects, whereby network size is particularly important. In general, goods with larger networks are more attractive for consumers, and thus, represent a competitive advantage. This principle can occur in two forms - positive and negative \cite{shapiro_information_1999}. In addition, network effects can be distinguished as direct and indirect \cite{reillier_platform_2017,shapiro_information_1999,clement_internet-okonomie_2016,eisenmann_opening_2008,gawer_industry_2014}.

In the following, both direct and indirect  network effects are briefly described  as they are corresponding to the underlying type of system.

\subsubsection{Direct Network Effects}

Direct or one-sided network effects occur when the number of users of a given good determines its value. Accordingly, the incentive to adopt a certain good  grows  with the number of users \cite{klemperer_network_2005}.With direct network effects, the benefit (\(U\)) that a user (\(n\)) gets from a network (\(U_{n}\)) depends not only on the (technical) characteristics (\(TK\)), but also on the total number of users (\(N\)) using the same network \cite{clement_internet-okonomie_2016}:

\begin{center}
\(U_{n}\) = \(U_{n}\)(\(N\), \(TK\)) with \(U_{n}\)(\(N\), \(TK\)) $<$ \(U_{n}\)(\(N^*\), \(TK\)) for \(N\) $<$ \(N^*\)
\end{center}

In this regard, the growth and development of networks can be described by "laws" \cite{clement_internet-okonomie_2016}. The most prominent one is Metcalfe's law \cite{metcalfe_law_1995}, more recent definitions come from Reed \cite{davis_power_2007} or Odlyzko \cite{briscoe_metcalfes_2006}.

With positive network effects, each new network participant increases the total value of a network by square or even more exponentially: each additional user increases the number of possible connections, thus, increasing the number of potentially reachable users. Consequently, the value of a network given to a customer increases disproportionately when a new participant enters the same network \cite{reillier_platform_2017,clement_internet-okonomie_2016}. New network participants increase the benefit of the already connected participants and at the same time make the network more attractive for further participants \cite{clement_internet-okonomie_2016,klemperer_network_2005}. In other words, the more people are connected within a network, the more valuable the network becomes for each participant who is part of it. If a new user wants to join the network, it benefits directly from other people whom it may want to reach \cite{reillier_platform_2017}. Economists call this a positive direct network effect, because one person has an influence on others, those who are new to the network as much as those who are already using it. Hereby, the increased value for uninvolved third parties is not or only partially compensated. A potential compensation would be, for example, if each new network participant would receive a (monetary) reward for the value added  by the existing and/or future participants.

Negative network effects occur, for example, when a telephone network is overloaded and no more telephone calls can be made. As a result, the value for each network participant decreases. A telephone network no longer has the function of connecting its network participants.

\begin{wrapfigure}{r}{0.5\textwidth}
    \begin{centering}
    \includegraphics[width=0.9\linewidth]{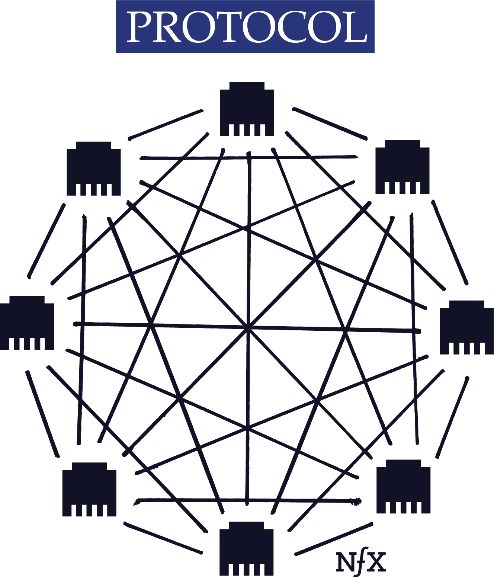}
    \par\end{centering}
    \caption{\label{fig:pne}Protocol Network Effects \cite{currier_network_2018}}
\end{wrapfigure}

Usually, direct or one-sided network effects result from an underlying protocol whereupon the effect is based \cite{currier_network_2018}. In this context, direct means that there is an immediate impact and increase of the value to its users. An example could be a protocol network effect (PNE) that directly connects consumers based on a standard definition. Hereby, both customer groups are on the same side and type. A well-known example for a one-sided network effect following this principle is \textit{Ethernet} or \textit{TCP/IP}. In addition, the above described landline telephone example by \cite{rohlfs_comunication_1974} represents a direct network effect based on PNE.

Figure \ref{fig:pne} illustrates how participants could be directly connected to each other using a network of a defined standard protocol.

\subsubsection{Indirect Network Effects}

The value for a customer in markets with network effects is typically determined by the number of customers on each side \cite{hagiu_strategic_2015,hagiu_information_2014}. If a network effect is indirect or two-sided, it is complementary to a related market: The benefit of a particular good depends on the dissemination of other goods. The networks value does not necessarily arise from a direct relationship between its participants, but from complementary goods within the entire system \cite{clement_internet-okonomie_2016}. Producers of a good benefit when other producers distribute their goods via the same network, making the offering of a platform particularly comprehensive. In this case, two sides of market actors come together through the introduction of an intermediary (e.g. a platform). Bilateral or multilateral markets arise in which the benefits of one side of the market is influenced by the other side. A relationship between these sides can be described as a platform?s attractiveness for the demand side and is strongly correlated with the offering of products or services from the supply side \cite{hagiu_marketplace_2015,armstrong_competition_2006,clement_internet-okonomie_2016,hagiu_strategic_2015}.

Accordingly, each user group has different reasons for participating in the network (e.g. buying and selling). By interacting, they create complementary value for each other \cite{hagiu_marketplace_2015}. For example, a seller derives more value from a network when more buyers are using the same network and \textit{vice versa} \cite{hagiu_strategic_2015}.

\subsection{Differentiation of Platforms}

As the economists Jean-Charles Rochet and French Nobel Prize Laureate Jean Tirole have already stated in their pioneering paper\footnote{A non-technical discussion of this paper and other early economic papers on platforms is given by \cite{schmalensee_platform_2014}.} on the general significance of network effects, "[m]any, if not most markets with network externalities are characterized by the presence of two distinct sides whose ultimate benefit stems from interacting through a common platform." \cite{rochet_platform_2003}. In this case, platform users of one customer group or side are influenced positively or negatively by platform users of the other side \cite{rochet_platform_2003}.

\begin{wrapfigure}{r}{0.5\textwidth}
    \begin{centering}
    \includegraphics[width=0.9\linewidth]{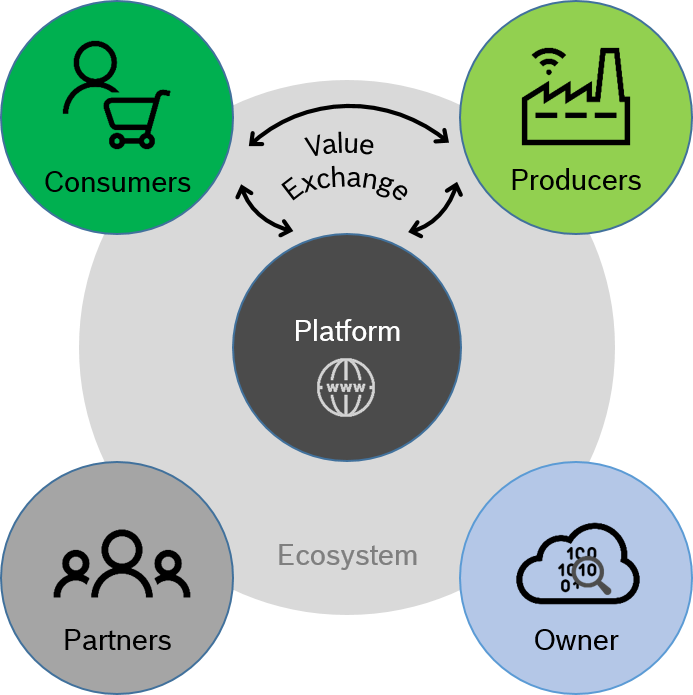}
    \par\end{centering}
    \caption{\label{fig:platform-ecosystem}Platform Ecosystems (own visualization, inspired by \cite{walter_4_major_2016})}
\end{wrapfigure}

To achieve this, platform operators must solve the so-called "chicken and egg" problem\footnote{Literally speaking, this problem includes that you can't have chickens without eggs, but you need chickens to get eggs \cite{evans_2016}.}, meaning that the value of each side is dependent on the adequate platform-use of the other side \cite{evans_2016,hagiu_strategic_2015}. As a result, buyers will not join a platform if its respective network does not provide enough sellers. Neither will sellers join if the network if it does not provide enough buyers. \cite{evans_2016,reillier_platform_2017}

Platforms are usually embedded within broader ecosystems of all the people, businesses, governments, regulations and other institutions that, 
as they interact with each other, affect the value a platform can create \cite{evans_2016,gawer_industry_2014,schreieck_governing_2011}.

In Figure \ref{fig:platform-ecosystem}, a platform ecosystem with consumers and producers, a platform owner as operator and platform partners is illustrated.

All actors interacting in the platform ecosystem are generally referred to as \textit{sides} \cite{schreieck_governing_2011}. Given the variety of actors involved, these platforms could be described as two-sided or multi-sided markets \cite{currier_network_2018,evans_platform_2011}. Accordingly, platform owners have to decide how many customer groups they want to connect and therefore how many sides to have \cite{evans_2016,hagiu_strategic_2015}.

\subsubsection{Two-Sided Platforms}

By bringing interests together, two-sided platforms provide a foundation for products \cite{currier_network_2018,evans_platform_2011}. Prominent examples are operating systems such as \textit{Linux}, \textit{Microsoft Windows}, \textit{Android} or \textit{Apple iOS} (excluding the app store as it would be a marketplace, see chapter \ref{subsubsec:marketplaces}). Each functionality (beside basic functions) is developed by the supply side and made available exclusively on the respective platform. Accordingly, these additional functionalities represent a function of the platform itself and cannot be considered independently \cite{currier_network_2018}.

Furthermore, the characteristics and advantages of the platform itself can provide compelling arguments for the benefits of a platform in relation to the network. For example, people buy smartphones which include an operating system, the design, technical features and performance of the phone as well as the app ecosystem. Accordingly, the product itself represents value, regardless of the network. In order to achieve the widest possible distribution of products on the supply side the platform must attract consumers, and hence, invest in marketing and sales \cite{currier_network_2018}.

\begin{wrapfigure}{r}{0.5\textwidth}
    \begin{centering}
    \includegraphics[width=0.9\linewidth]{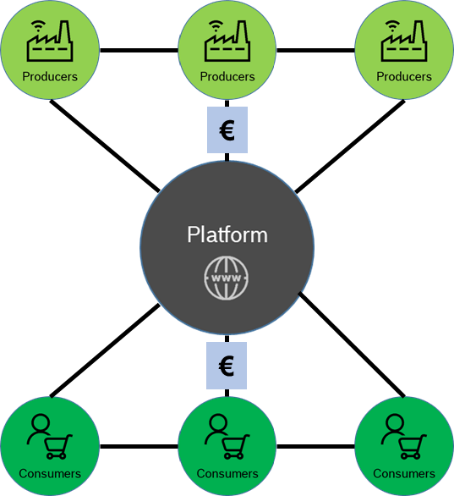}
    \par\end{centering}
    \caption{\label{fig:platform-2-sided}Two-sided Platform (own visualization, inspired by \cite{currier_network_2018})}
\end{wrapfigure}

Figure \ref{fig:platform-2-sided} illustrates the role of the platform as a central intermediary, bringing together two groups of participants. For example, users of a dedicated platform gain more value when several other users utilize the same platform because they can interact more easily.

We assume that standard IoT-platforms\footnote{For a more detailed discussion about platforms in industrial sphere, see \cite{gawer_industry_2014}.} are also an example for two-sided platforms as they are comparable to an operating system. Platform operators provide an infrastructure of hardware and software, enabling services to be offered, and thereby form a business ecosystem and potentially generate network effects \cite{gawer_industry_2014,schreieck_governing_2011}. Users and devices represent the two sides of the IoT-platform. Value-adding services are only provided by the platform operator. Complementary companies have no access to the platform \cite{schreieck_governing_2011}.

Along the horizontal value chain, corresponding platforms such as \textit{AWS}, \textit{Microsoft Azure} and \textit{Siemens Mindsphere} offer cloud-based (and on-premise) solutions for consumers, on which vertical applications could be built upon in every IoT-domain \cite{schreieck_governing_2011}. In this way, they enable a connectivity standard on which IoT services could be developed.
If standard IoT platforms also include a marketplace, they can be described as advanced IoT platforms. Hereby, platform operators open their ecosystem to third parties, allow partners to interact directly with the demand-side and offer consumers additional services, e.g. new applications similar to \textit{iOS} \cite{schreieck_governing_2011}. However, \cite{schreieck_governing_2011} state that no advanced IoT platform functionality has yet been able to establish itself. As one potential cause, we would like to argue for the lack of consistent, domain-specific standards, which could rapidly trigger direct network effects.

\subsubsection{\label{subsubsec:marketplaces}Marketplaces as Multi-Sided Platforms}

Marketplaces bring together different customer groups together and primarily create value by enabling direct transactions between two or more customer groups from multiple sides. In their function as intermediary matchmakers, they facilitate access to customers. They provide the necessary infrastructure for interacting and enable value-creating activities between at least two actors, usually producers and consumers \cite{reillier_platform_2017,hagiu_marketplace_2015,armstrong_competition_2006,hagiu_platforms_2015,gawer_industry_2014,hagiu_strategic_2015,taeuscher_understanding_2017,parker_platform_2016}.
Therefore, matchmakers usually operate a physical or digital place where members of these different groups get together \cite{evans_2016,reillier_platform_2017,kollmann_virtual_1999}.

\begin{wrapfigure}{r}{0.5\textwidth}
    \begin{centering}
    \includegraphics[width=0.9\linewidth]{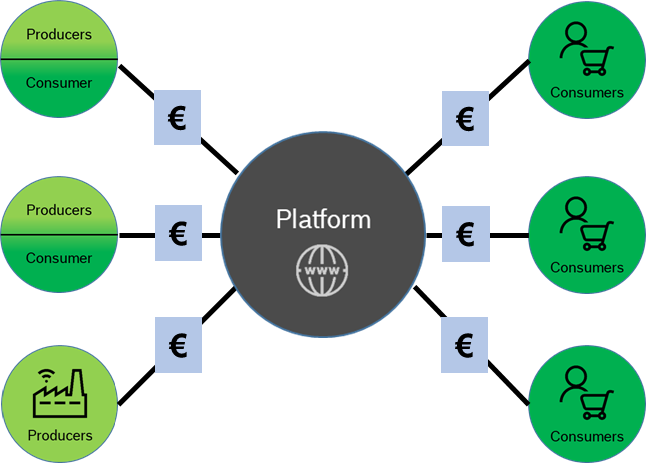}
    \par\end{centering}
    \caption{\label{fig:marketplace}Marketplace (own visualization, inspired by \cite{currier_network_2018})}
\end{wrapfigure}

A digital marketplace is a virtual market space that attracts users and producers alike, brings them together and allows digital business transactions to be conducted as shown in Figure \ref{fig:marketplace}. Electronic marketplaces that use information technology can be named digital marketplaces. In other words, they are providing trading systems for specific business transactions within information or communication networks \cite{clement_internet-okonomie_2016}. They focus on those participants who want to offer a product or service (producers) and those who want to buy this product or service (consumers).

Under predetermined rules, value-creating activities operated by marketplace owners involve an exchange of information, goods or services and a payment mechanism \cite{parker_platform_2016}.

This connection can be illustrated using the example of \textit{Uber}: As soon as a potential customer opens the smartphone app, the availability and location of potential drivers is transmitted (information). If customer and driver come together, a service is provided in the form of a journey from A to B (service). In return, the customer pays the driver a monetary amount (payment), which is calculated according to certain parameters.

Hereby, the marketplace operator fulfils an overview function of the market situation and coordinates economies of scale. By means of a transaction platform, it brings together supply and demand, both qualitatively and quantitatively \cite{reillier_platform_2017,taeuscher_understanding_2017,kollmann_virtual_1999}.

Thereby, a functional distinction can be made between technical or business-related elements. Technical elements include search functions or encryption; business elements include requests for quotations, catalog and matching systems or auctions \cite{clement_internet-okonomie_2016,parker_platform_2016,kollmann_virtual_1999}.

Exemplary for multi-sided transaction platforms are  marketplaces like \textit{Amazon Marketplace} or well-known examples from the sharing economy like \textit{Uber}, \textit{Airbnb} and \textit{eBay}.

Whereas the value of a two-sided platform includes the functions provided, the value of a marketplace consists mainly of its network size. The more users are participating in the marketplace, the more attractive it is. \textit{eBay} is a good example of a marketplace consisting mainly of network size rather than features, as after 16 years it consists mostly of the same functionality that was provided originally \cite{currier_network_2018}.

Likewise, network effects play a decisive role in the attractiveness of \textit{Matchmaker}-platforms, that can be described as databases for a large number of goods available for sale. Access to these goods is granted via the respective websites. The platform derives more value to the buyers when there are more sellers and \textit{vice versa}. Anyone who has something to sell will benefit from using the marketplace, as buyers will follow sellers, and if all possible buyers are in the same place, this should maximize profit for all sellers.

\subsection{\label{subsec:impact_network_effects}Impact of Network Effects}

Network effects generally increase the value provided by the platform for each individual participant in a respective network. As the number of users increases, the platforms become more attractive. Supply and demand stimulate each other. Through strong, positive, indirect network effects, platforms can quickly scale their business model. Positive economies of scale result in a larger number of units produced or sold, which is associated with decreasing costs per unit. Once a critical mass is reached, the platform can establish itself as a \textit{de facto} standard and dominate in its market \cite{evans_2016,reillier_platform_2017,shapiro_information_1999,clement_internet-okonomie_2016}. This effect can be seen as extremely beneficial and profitable for platform operators\footnote{For further information regarding profit of platform operators see \url{https://www.applicoinc.com/blog/platform-vs-linear-business-models-101/}}.

Especially in the case of digital goods, in which the costs for production are mostly fixed and the costs for the production of an additional unit are negligible \cite{shapiro_information_1999,clement_internet-okonomie_2016}, such falling marginal costs tend to be the rule. In contrast to bilateral trade, which is characterised by high transaction costs, platforms can reduce these costs and communicate information on supply and demand at low costs \cite{hagiu_multi_platforms_2006}. Although the development and maintenance of the infrastructure (\textit{platform curation}) of platforms is expensive, additional users hardly cause additional costs \cite{hagiu_strategic_2015,schmueck_democratization_2019}.

Apart from advantages, especially with respect to the platform operators, network effects can also cause economic imbalances\footnote{Economic imbalances are caused by "spillover effects", where costs or benefits of an economic interaction have to be incurred by parties that are not directly involved in the interaction\cite{klemperer_network_2005,herd_towards_2018}.} \cite{herd_towards_2018}, leading to undesirable economic consequences. For example, these may include inefficient and unstable equilibria and path dependencies, resulting in strong "\textbf{winner-takes-it-all-dynamics}" \cite{schmueck_democratization_2019,herd_towards_2018}. The successes on these markets follow a distribution that is reminiscent of mathematical power laws: In order to ignite, platforms endeavour to reach a critical mass\footnote{Strategies to secure a critical mass are discussed in \cite{evans_2016,eisenmann_opening_2008,eisenmann_strategies_2006}.} of users that will encourage users of other systems to switch \cite{evans_2016,shapiro_information_1999}. As a result of a positive feedback cycle, the strong will get stronger and the weak will get weaker\cite{vanalstyne_pipelines_2016,shapiro_information_1999,clement_internet-okonomie_2016}.

If strong network effects occur in markets, users can be tied to a specific provider (\textbf{lock-in effects}) -- even if the given provider offers inferior products compared to others. The lock-in consists of significant switching costs, if customers want to switch from one platform, system or technology to another \cite{shapiro_information_1999,clement_internet-okonomie_2016,klemperer_network_2005}. Managing these costs is very difficult for all parties involved, as they are usually not obvious and occur at organizational and technical level. The expenses can either be material (e.g. access costs, participation and registration fees) or immaterial (e.g. time expenditure, specific learning costs). In the case of immaterial expenses, costs may arise for staff training as well as the technical and organisational integration into existing systems. The users have learned how to work with legacy systems and got used to defined user interfaces \cite{clement_internet-okonomie_2016}. Software may also have been used to create files with non-migratable data and, if necessary, auxiliary programs may have been created for their use \cite{clement_internet-okonomie_2016}. In addition, material lock-in effects can also be caused by strategic decisions of the operating platform providers. Common types are rewards for repeated purchases or pricing strategies that initially attract new customers to become part of the network and subsequently increased utilization prices over time \cite{shapiro_information_1999,eisenmann_strategies_2006,clement_internet-okonomie_2016,hagiu_strategic_2015}.

\begin{wrapfigure}{r}{0.5\textwidth}
    \begin{centering}
    \includegraphics[width=0.9\linewidth]{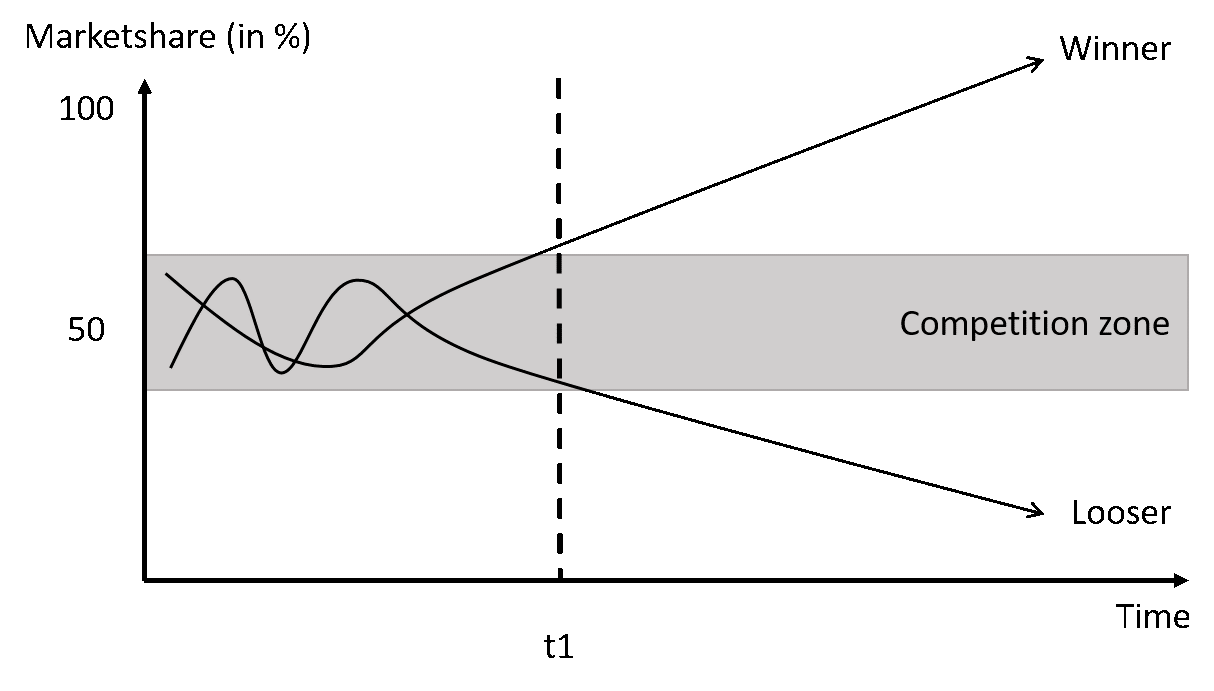}
    \par\end{centering}
    \caption{\label{fig:winner}Winner-takes-all-markets \cite{shapiro_information_1999,clement_internet-okonomie_2016}}
\end{wrapfigure}

Described effects and \textbf{monopolization tendencies} can be observed in electronic markets. Although two companies have the same resources and offer a good that provides value to consumers, one company becomes dominant after a certain time as shown in Figure \ref{fig:winner}. This tilting behavior is caused by positive self-reinforcing effects  and leads to a widening of the gap between the market participants: While one company benefits over-proportionally from the market growth, other companies lose more and more importance \cite{armstrong_competition_2006,shapiro_information_1999,clement_internet-okonomie_2016}. In addition, \cite{herd_towards_2018} also observed this tilting behavior. In an agent-based modeling and empirical game theory, they simulated that one platform always dominated the market and monopolized centralization tendencies arise naturally.

\begin{wrapfigure}{r}{0.5\textwidth}
    \begin{centering}
    \includegraphics[width=0.9\linewidth]{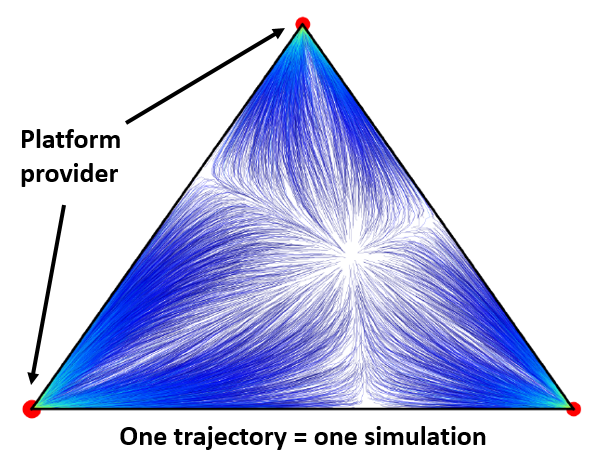}
    \par\end{centering}
    \caption{\label{fig:monopolistic}Monopolistic tendencies in platform markets\cite{herd_towards_2018}}
\end{wrapfigure}

Figure \ref{fig:monopolistic} shows a model with multiple simulation runs, in which multiple users were greedily optimizing their platform behavior. Each blue filament represents a trajectory of the simulation. Each trajectory is a representation of the distribution of users among the three different platforms. The distance to the respective platform provider is inverse to the number of users on the platform (more users = smaller distance). Since the platform at the bottom left of Figure \ref{fig:monopolistic} has the best cost function, it "wins" most often (higher probability). In this simulaton, it is not possible to predict which provider will win.

Even though the providers were allowed to switch between three platforms, the result was not dependent on starting conditions: one random platform was dominating and a monopoly arised. Accordingly, the success of a platform -- excluding marketing effects\footnote{Marketing is crucial for market and platform growth, as user expectations are critical in order to become a standard or at least achieve a critical mass. As \cite{shapiro_information_1999} pointed out, "the product that is expected to become the standard will become the standard". Resulting of those self-fulfilling expectations are a manifestation of positive-feedback economics and 'bandwagon effects \cite{shapiro_information_1999}.} -- cannot be planned. After the dominant position is established, it is quite hard for others to enter the market \cite{herd_towards_2018,choudary_dangers_2017}. This can also be observed in Figure \ref{fig:monopolistic}, as there is no direct line from the top and bottom right to bottom left, if market shares are already established.

While creating barriers to entry, established platform operators occupy privileged and often hard-to-assail positions as the eternal intermediary in their respective markets \cite{hagiu_strategic_2015}. Moreover, their recommendation algorithms\footnote{For a general discussion of the use of search and matching algorithms to facilitate transactions in online markets, see \cite{einav_peer_markets_2015,clement_internet-okonomie_2016}.} are based, at least partially, on analysis of a very large number of user interactions. Accordingly, the quality and benefit of the service depends on its number of users.

Although the platform offers added value for every user, it is the platform operator who benefits the most. Being the monopolistic player, they benefit from disproportionately high margins and rapid growth \cite{schmueck_democratization_2019}. As a result, everybody wants to own the respective platform and nobody wants to be locked in on other platforms. In business IoT platforms, this deadlock  results in small platforms without benefits of scaling networks.

\section{\label{sec:marketplaces}Marketplaces as intermediary coordinator}

The second chapter introduced the basics of network effects and showed a classification that correlates with system types such as a two-sided platform and a marketplace. Continuing, we want to focus on marketplaces as they are relevant for the coordination of goods or services between different units, having an economic background in mind. Hereby, digital networks share many characteristics with real networks, such as communication and transport networks \cite{shapiro_information_1999}. They are not dependent on physical locations and national borders or specific hardware and operating systems. They rather link and connect users in a digital way by providing services based on information technologies.

\subsection{\label{subsec:coordination_function}Coordination Function}

Marketplaces enable digital ecosystems and support a multilateral connection of business partners. This results in opportunities that are particularly interesting for companies that want to develop new business models and markets beyond established industries.

By creating a \textbf{logically central coordination point}, transaction costs can be reduced (see Baligh Richartz effect \cite{baligh_vertical_market_1964}): Supply and demand come together at a single point, increase market transparency and reduce search costs \cite{clement_internet-okonomie_2016,hagiu_strategic_2015,taeuscher_understanding_2017,hagiu_multi_platforms_2006}. By providing a place where offers can be bundled \cite{kollmann_virtual_1999}, marketplaces remove local or time restrictions and offer high scalability \cite{kollmann_virtual_1999,kollmann_elektronische_1998}. Information technologies enable simultaneous coordination of activities, thereby reducing transaction costs and minimizing misallocations. By gaining access to physically inaccessible markets and through the development of new market services (e.g. digital services for existing products), companies can generate increasing revenues that would not be possible without marketplaces. These services can be provided directly to the consumer and, due to their digital character, are highly scalable.

Altogether, a marketplace represents an effective trading environment with efficient market co-ordination \cite{kollmann_virtual_1999}.

\begin{figure}[tb]
    \begin{centering}
    \includegraphics[width=1\textwidth]{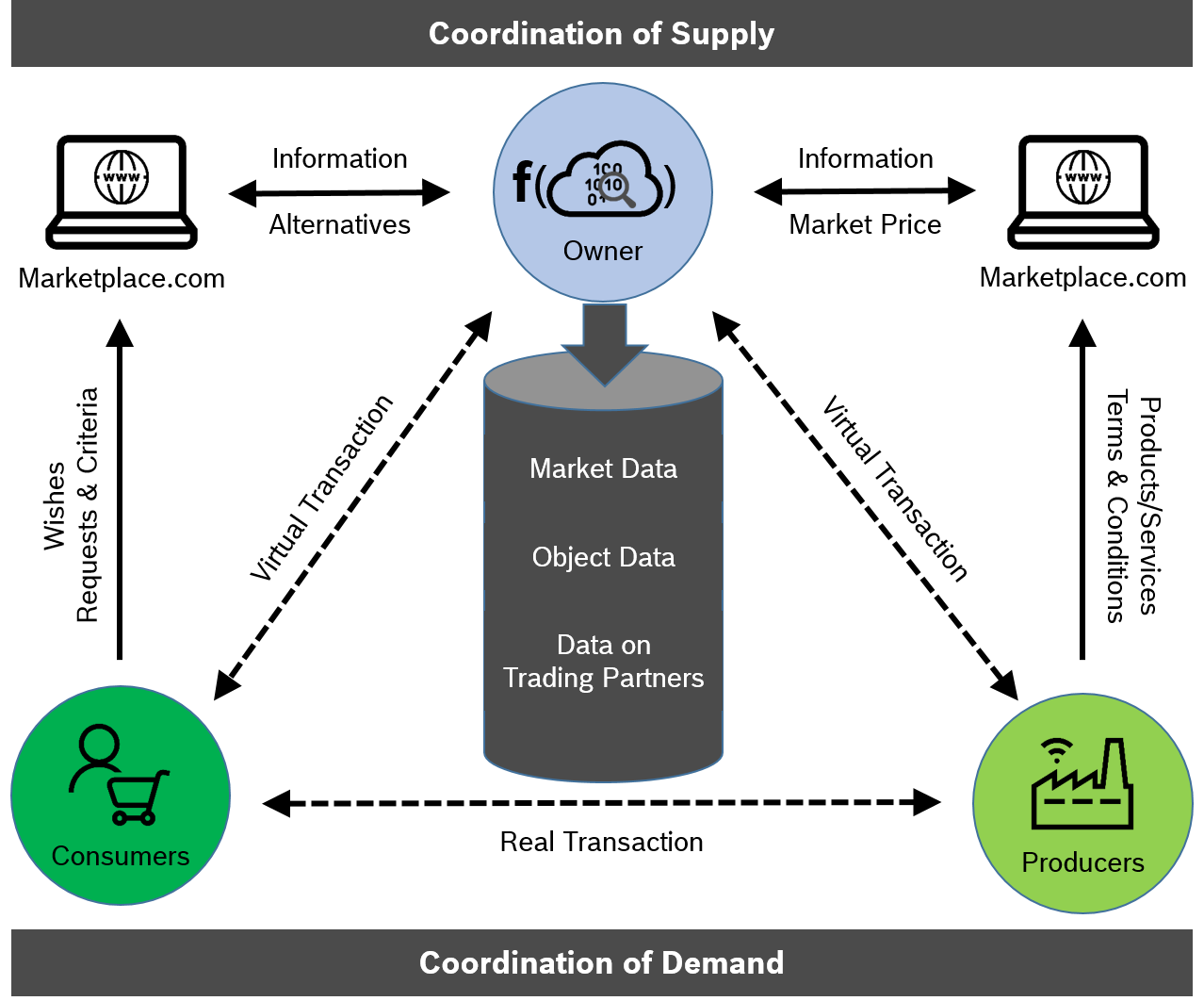}
    \par\end{centering}
    \caption{\label{fig:marketplace-function}Flow of values and functions in marketplaces (own visualization, inspired by \cite{kollmann_virtual_1999})}
\end{figure}

In contrast to online shops where only a trading environment is provided (e.g. in the form of lists which the inquirer has to browse through and request independently), digital marketplaces also offer an active coordination service. As part of the so-called matchmaking between supply and demand, the marketplace operator registers the object-related supply and demand before actively suggesting potential transaction partners to marketplace participants \cite{kollmann_virtual_1999}. The added value -- for which the operator is ultimately paid for by the market participants -- consists on one hand of achieving market transparency for the demander and on the other hand in a more efficient brokerage of objects for the provider \cite{kollmann_virtual_1999}.

This \textbf{matching function \textit{f(x)}}, illustrated in Figure \ref{fig:marketplace-function} at the owner instance, is crucial for marketplace operators when it comes to competition between rival electronic marketplaces. This function  establishes the connection between quantitative and qualitative competitive features and can be considered as a central control mechanism\cite{kollmann_matching_2005}. In order to achieve the best result between supply and demand, the matching function obtains the entire market knowledge as input. The more inputs are available for this function, the more valuable are the results for the consumer. Consequently, it is more attractive for consumers to use this function if they get better results.

\subsection{Drawbacks establishing a centralized intermediary}

Currently, the majority of virtual marketplaces are managed and operated by a single legal entity. Consequently, the entire transaction process of finding, filtering and communicating is provided by a single organization \cite{shapiro_information_1999}. Hereby, the marketplace operator acts as an intermediary, collecting all available information on the transaction process \cite{kollmann_toward_2019}. Considering the intermediary position of the marketplace operator as a central authority, not only advantages but also various disadvantages for sellers and buyers as well as the overall market situation can be identified. 

In this context, the power of the identity-creating intermediary is enormous. By deliberately avoiding interoperability, lock-in effects  are caused or reinforced by a technical standard that makes it difficult or costly to use multiple platforms (\textit{"multihoming"}, \cite{rochet_platform_2003,reillier_platform_2017,armstrong_competition_2006,gawer_industry_2014}). Single Sign-On solutions are very attractive for users, but often the data is not transferable between different identity providers.

Matchmakers also collect considerable amounts of data about their customers and obtain transparent knowledge of the market situation. This includes (transaction) metadata which provides platform operators with additional knowledge about the interaction partners and represents an immaterial economic value. By collecting, analyzing and combining the data, platform operators can build up an enormous amount of knowledge about the interaction partners, which they are only able to do in their role as intermediaries. 
By aggregating their knowledge, platforms can gain a competitive advantage from which they can benefit in various ways. With every business transaction, platform operators have more data at their disposal, which they can analyze  and use to develop their own offers in particularly lucrative markets. This not only allows them to follow the general price development over time, but also provides them with detailed information on each transaction concluded. Information such as price, quantity and time are of interest to sellers, as they can use this information to evaluate other competitors. Through metadata analysis, platforms can e.g. identify particularly successful start-ups and buy them if they are interested.

It is remarkable that marketplace owner companies rarely contribute themselves to creating value on the supply side. They provide no services, only their marketplace platform. Their profit increases exponentially depending on the number of transactions offered and the marginal costs are decreasing accordingly. This leads to an imbalance of value distribution within ecosystems where marketplace users create value while marketplace owners benefit the most \cite{kollmann_e-business_2019}. \textit{Uber}, in this context, retains up to 28.5\% of the travel costs, yet having neither cars nor drivers \cite{schmueck_democratization_2019}.

Producers on the platform have no significant bargaining power: either they submit to the conditions of the platform operator -- from high revenue shares to content requirements -- or they have no access to consumers using the platform.

At the same time, market-dominating marketplace operators can expand their business model and penetrate the upstream and downstream value chain of traditional companies. In this context, there is a risk that marketplace operators themselves could become competitors and, with the help of the data collected and their market dominance, could crowd out smaller suppliers. For their part, the operators could build up a closed ecosystem of hardware, software, services and content and, in the process, purchase assets from the traditional business. By entering the market, they could then act as producers or service operators, and thus, gradually replace the traditional market participants. This scenario can already be seen in the B2C sector with \textit{Amazon}. The company acts as a platform operator with its marketplace. In addition, \textit{Amazon} also sells its own products via this marketplace (e.g. \textit{Amazon Basic}) and will in future act as a logistics service provider itself (\textit{Prime Air}). Since \textit{Amazon} collects all metadata of the transactions, they gain knowledge on interesting goods with high sales figures and the current price. On one hand, \textit{Amazon} can sell goods at a more attractive price \cite{brandom_monopoly_2018}. On the other hand, it is also able to place these goods at the top of the search, as they are providing this function and can manipulate the information \cite{shapiro_information_1999}. 

Another possibility for the marketplace operator is to establish personal prizing\cite{shapiro_information_1999}. Compared to dynamic pricing where different prices for the same product are available at different times, personal pricing creates different prices for different types of consumers. This is possible because the platform collects not only transaction-related data but also consumer-related data. This creates an information asymmetry\footnote{Asymmetric information derives from market theory considerations of the Nobel Prize Laureates Stiglitz, Akerlof and Spence and can be described with the principal-agent-theory. For further information, see \cite{clement_internet-okonomie_2016,akerlof_lemons_1970}.} where the consumer does not know what kind of information is available to the marketplace operator. This leads to an environment where consumers believe that they see the real market price even if they do not \cite{walker_how_2017}. A prominent example that shows this asymmetry of information is \textit{Uber}. They apply a so-called "route-based pricing", which is based on certain criteria, such as the type of credit card used (e.g. private or corporate). \textit{Uber} says that their pricing is based on their understanding of demand patterns rather than individual drivers \cite{gonzaga_personal_pricing_2018}. 

In this context, platform operators can also trigger a slow increase in transaction costs by creating a dependency of interaction processes and making it necessary for more and more processes to be carried out via the platform. Charing for each transaction, the platform has the pricing power.

As described in the context of the coordination function \textit{f(x)}, matchmakers enable consumers to provide and filter the right information. The use of such a function is necessary today, as access to information is no longer a problem arising from the use of digital technologies, but the overload of information. It is therefore beneficial for consumers to locate, filter and communicate anything useful to them \cite{shapiro_information_1999}. If the rules for filtering and providing information are not common knowledge and are only provided by a single entity, that entity has the power to change and modify rules in order to achieve objectives that may conflict with general social welfare. For example, \textit{Trivago} manipulated search result lists by showing not the best price for a booking at the top, but the results from which \textit{Trivago} could earn the most revenue \cite{duran_trivago_2020}. Another prominent example of non-obvious rules is the court case against the rating platform \textit{Yelp}, on which positive ratings were sorted out by an algorithm \cite{hempel_yelp_2020}.

Overall, an established platform monopolist can use the mechanisms described above to increase its profits and enhance incentives for rent-seeking behavior \cite{gonzaga_personal_pricing_2018}. Consequently, the fundamental question can be asked whether platform monopolies, including negative side effects, are an unavoidable part of today's economy or whether there is a solution that obtains the advantages of network effects on the basis of fair mechanisms.

\section{Fair \& Open Marketplaces}

While Chapter \ref{sec:network_effects} introduced the general understanding of network effects and platforms, Chapter \ref{sec:marketplaces} discussed how a marketplace works and suggested possible disadvantages of a centralized design. These are contrary to the principle of equal participation and lead to an unbalanced distribution of values between operator and participant of the ecosystem. In this chapter, we want to outline how a possible solution can be achieved, overcoming information asymmetries and building a fair system in which each of the participants is able to maximize its own value by avoiding a centralized platform operator with super-ordinary profits.

\subsection{Coopetition as Nucleus}

Given the principles of network and platform economics, it is extremely difficult to establish a successful platform in the market. Besides solving the chicken-and-egg problem, companies are mainly confronted with strong competition and fiercely competitive environments. We assume that market participants who put themselves in a position where they can connect their services with others, create combined offers and sell them successfully on the market, most likely survive. Although this form of inter-company cooperation questions the competitive logic of entrepreneurial action, services of different providers can be combined to provide a highly competitive offering. As possible solutions, \cite{shapiro_information_1999} introduced alliances and open standards, both solving different problems.

Open standards solve the acceptance and technological risk that consumers are facing. It is not enough to have an open standard. Direct competitors must be able to compete based on this standard \cite{shapiro_information_1999}. This combination will increase acceptance by improving compatibility, interoperability and future competition.  

As described in the Chapter \ref{sec:marketplaces}, the coordination function is the core element of the matchmaker. Duplicating this function to several instances is not a solution, since it consists of network effects as multiple platforms tend to have winner-takes-it-all-dynamics (see chapter \ref{subsec:impact_network_effects}). The more input is available from the participants, the better the result becomes. The better the result, the higher the participant's value (see chapter \ref{subsec:coordination_function}).

If this \textbf{coordination function} is combined with \textbf{alliances}, a cooperative system can be established that \textbf{operates the logically central coordination function}, but is operated and governed in a decentralized way by several different partners. As a result, possible information advantages for one party can be prevented.

\begin{wrapfigure}{r}{0.5\textwidth}
    \begin{centering}
    \includegraphics[width=0.9\linewidth]{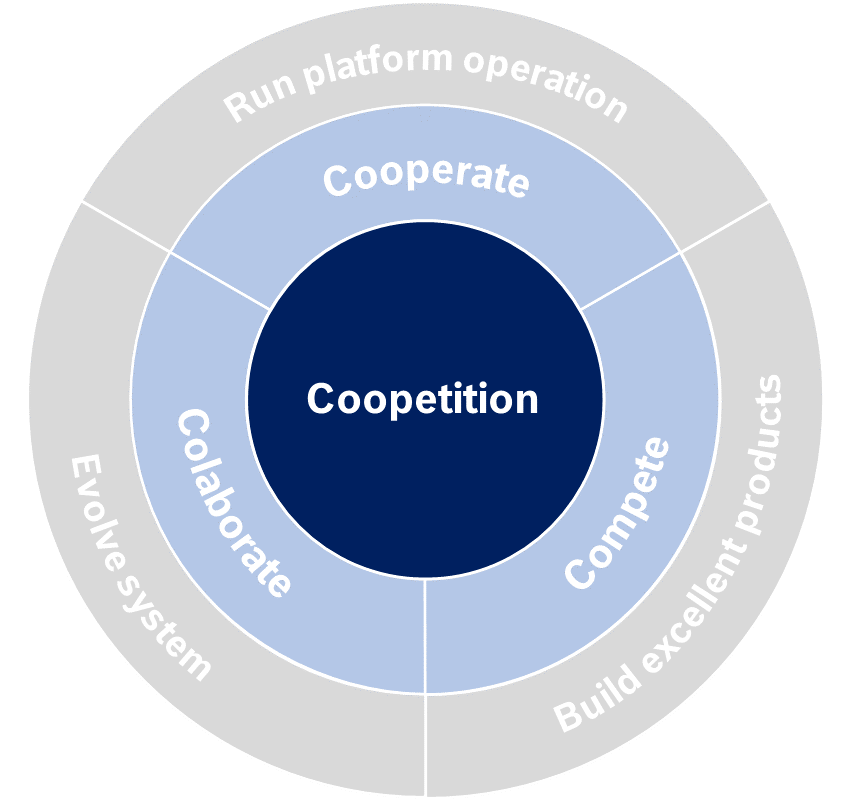}
    \par\end{centering}
    \caption{\label{fig:coopetition}Coopetition}
\end{wrapfigure}

In form of so-called \textit{\textbf{coopetition}}, visualized in Figure \ref{fig:coopetition}, we see a high potential to cooperate with competitors in a collaborative ecosystem. Resources and services can be interconnected between participants and mutually utilized. Specifically, we define coopetition as a combination of cooperation, collaboration   and competition. 
Companies \textbf{cooperate} and combine their resources to develop and operate the technical system or platform, without a single company controlling the network. They \textbf{collaborate} in the further development of the platform. Participating on the platform, they \textbf{compete} with each other at the level of specific products and services. Here, the link to the customer is established by individual companies (or associations) to create an attractive and competitive market offer. We assume this results in added value for all those involved in the ecosystem and competitive advantages due to coopetition. Monopolies by individual companies, as observed in the current platform economy, can be avoided.

As a substantial criterion for success we believe that a common, superordinate purpose within the cooperation, based on which commonly applicable standards, principles and rules of coopetitionare are defined, is essential. Once there is a mutually accepted consensus, specific goals can be realized based on it, allowing collective response to market changes. As a result, ecosystem participants have the opportunity to combine their services in a cost-effective and market-attractive manner. 

\subsection{Economy of Things as fair coopetition-environment}

As part of digitization and the resulting requirements, the \textit{Internet of Things} (IoT) or \textit{Web2.0} emerged. Various technologies, such as the semantic web or distributed databases, enable connections between cross-domain things and edge devices. IoT aims to enable connectivity between devices and automate processes.

This concept can be further developed to an \textit{Economy of Things} (EoT) within \textit{Web3.0}, where not only things are connected but also economic interactions between those become feasible. Those systems will increasingly exhibit the following characteristics:

\begin{center}
    \begin{tabular}{ p{0.20\linewidth} p{0.70\linewidth} } 
        \textbf{Dynamism:} & Relations can form and break up dynamically.  \\
        \textbf{Heterogeneity:}  & Participants differ with respect to characteristics and behavior. \\
        \textbf{Autonomy:} & Participants act autonomously. \\
        \textbf{Self-interest:} & Upon the cooperation, participants pursue their own goals.
    \end{tabular}
\end{center}

EoT may be based on a decentralized infrastructure, where each company determines which information is shared with which business partner when releasing resources. In contrast to centralized structures, where power, control and trust is consolidated to the greatest possible extent, a decentralized concept distributes responsibilities within the community. Power and information imbalances, unlike in centralized structures, can be avoided. The platform's network effects will be beneficial for the entire system and not for the disproportionately high benefit of individual actors who take over the orchestration. 

To create such an EoT ecosystem, several elements need to be addressed. These have to meet the coordination mechanisms described in chapter \ref{sec:marketplaces}. In order to solve some of the disadvantages described, we propose a combination of the following main building blocks:

\begin{center}
    \begin{tabular}{ p{0.40\linewidth} p{0.50\linewidth} } 
    \textbf{Organization and Governance:} & Trustworthy and open 3\textsuperscript{rd} party organization with transparent and non-discriminating rules. \\ 
    \textbf{Technology and Operation:} & Decentralized and distributed technologies to avoid
    possible information asymmetries. \\ 
    \end{tabular}
\end{center}

At technological level, the approach of openness and cooperation has several implications. First, the system that delivers the results must use or establish openly available standards to foster widespread acceptance and adoption. In addition, operational aspects may be solved by using decentralized networks. Notable developments include decentralized technologies such as \textit{Blockchain} and \textit{Distributed Ledger Technologies} (DLT).

Moreover, the avoidance of information asymmetry and information surplus of one party must be achieved. However, it is crucial that the coordination aspects are maintained as a logical central point.  The simplest way to meet these requirements is to open up all information to everyone (like the cryptocurrency Bitcoin), so that all parties end up with the same information. Nonetheless, it is not sufficient to disclose all relevant information for each use case, as there may be requirements such as privacy concerns for dedicated input data. Consequently, privacy enhancing combinations of different types of cryptographic protocols (such as \textit{multi-party computation} (MPC), \textit{homomorphic encryption}, \textit{zero-knowledge proofs} (ZKP) and \textit{zkSNARKs}), could be used  to solve certain requirements. 

Furthermore, the development and establishment of such systems is a great effort and complicated from an organizational and technical point of view. In order to build up such a basis and realize all coordination aspects, it seems appropriate to combine forces following of the coopetition approach. To coordinate the building blocks in a focused and open manner, we propose several basic principles that are necessary to establish acceptance. These principles form the basis for a fair and neutral competitive environment, that focuses primarily on the interests of the platform users and enable the rapid scalability of the system:

\begin{center}
    \begin{tabular}{ p{0.20\linewidth} p{0.70\linewidth} } 
        \textbf{Openness:} & The cooperation is permanently open to all participants who comply
        with and wish to participate in the principles that apply to everyone. \\
        \textbf{Neutrality:} & No single ecosystem participant should dominate the cooperation. Power, control and resources are fairly distributed among the participants. In addition, the network effects of the established platform are aggregated for the common gain of all participants, which is in contrast to a centralized platform where the operator aggregates a super proportional fraction of the network gain. \\
        \textbf{Transparency:} & Transparent business model, organizational structures, regulations
        and decision-making processes. \\
        \textbf{Integrity:} & Unique identification of actors, authenticity of the information    exchanged while preserving privacy. \\
        \textbf{Sovereignty:} & Data sovereignty for the authorized party, non-discriminatory and
        legally admissible access to data for all market participants. No existential dependency of participants on the cooperation. \\
        \textbf{Sustainability:} & Flexible response to changes in the ecosystem with resilience in
        basic principles. \\
    \end{tabular}
\end{center}

It should be noted that EoT can only be as decentralized as its most centralized subsystem \cite{srinivasan_decentralization_2017,buterin_decentralization_2017}. Therefore, alignments to legal rules as well as various strategic decisions such as platform design including basic functionalities, governance and pricing structure have to be made. For a brief discussion, see \cite{vanalstyne_pipelines_2016,eisenmann_strategies_2006,hagiu_platforms_2015,hagiu_strategic_2015}.


\selectlanguage{american}%
\bibliographystyle{myunsrt}
\bibliography{mechanismsOfIntermediaryPlatforms}

\begin{thebibliography}{10}

\bibitem{statista_biggest_companies}
Statista:
\newblock Biggest companies in the world 2019
\newblock
  \url{https://www.statista.com/statistics/263264/top-companies-in-the-world-by-market-value/}.

\bibitem{evans_2016}
Daniel Evans and Richard Schmalensee:
\newblock {\em Matchmakers}
\newblock Harvard Business Review Press.

\bibitem{rochet_platform_2003}
Jean-Charles Rochet and Jean Tirole:
\newblock Platform Competition in two-sided Markets.
\newblock {\em European Economic Association\/}.

\bibitem{reillier_platform_2017}
Laure~Claire Reillier and Benoit Reillier:
\newblock {\em Platform Strategy}.

\bibitem{hagiu_marketplace_2015}
Andrei Hagiu and Julian Wright:
\newblock Marketplace or Reseller?
\newblock {\em Management Science\/}.

\bibitem{armstrong_competition_2006}
Mark Armstrong:
\newblock Competition in two-sided markets.
\newblock {\em The RAND Journal of Economics\/}.

\bibitem{statista_us_amazon_market_share}
Statista:
\newblock U.S. Amazon market share in e-commerce and retail 2021
\newblock
  \url{https://www.statista.com/statistics/788109/amazon-retail-market-share-usa/}.

\bibitem{bmwi_b2b_2019}
Christian Lerch, Niclas Meyer, Djerdj Horvat, Thomas Jackwerth-Rice, Angela
  J\"{a}ger, Michael Lobsinger, and Nadia Weidner:
\newblock Die volkswirtschaftliche Bedeutung von digitalen B2B-Plattformen im
  Verarbeitenden Gewerbe.
\newblock {\em Bundesministerium f\"{u}r Wirtschaft und Energie (BMWi)\/}.

\bibitem{vanalstyne_pipelines_2016}
Marshall~W. Van~Alstyne, Geoffrey~G. Parker, and Sangeet~Paul Choudary:
\newblock Pipelines, Platforms, and the New Rules of Strategy.
\newblock {\em Harvard Business Review\/}.

\bibitem{shapiro_information_1999}
Carl Shapiro and Hal~R. Varian:
\newblock {\em Information Rules - A Strategic Guide to the Network Economy}
\newblock Harvard Business School Press.

\bibitem{eisenmann_strategies_2006}
Thomas~R. Eisenmann, Geoffrey Parker, and Marshall Van~Alstyne:
\newblock Strategies for Two-Sided Markets.
\newblock {\em Harvard Business Review OnPoint\/}.

\bibitem{hagiu_platforms_2015}
Andrei Hagiu and Julian Wright:
\newblock Multi-sided platforms.
\newblock {\em International Journal of Industrial Organization\/}.

\bibitem{schmalensee_introduction_2011}
Richard Schmalensee:
\newblock Jeffrey Rohlfs' 1974 Model of Facebook: An Introduction.
\newblock {\em Competition Policy International\/}.

\bibitem{rohlfs_comunication_1974}
Jeffrey Rohlfs:
\newblock A Theory of Interdependent Demand for a Communications Service.
\newblock {\em The Bell Journal of Economics and Management Science\/}.

\bibitem{clement_internet-okonomie_2016}
Rainer Clement and Dirk Schreiber:
\newblock {\em Internet-\"{O}konomie}
\newblock Springer Gabler.

\bibitem{eisenmann_opening_2008}
Thomas~R. Eisenmann, Geoffrey Parker, and Marshall Van~Alstyne:
\newblock Opening Platforms: How, When and Why?
\newblock {\em Harvard Business School\/}.

\bibitem{gawer_industry_2014}
Annabelle Gawer and Michael~A. Cusumano:
\newblock Industry Platforms and Ecosystem Innovation.
\newblock {\em Journal of Production Innovation Management\/}.

\bibitem{klemperer_network_2005}
Paul Klemperer:
\newblock Network Effects and Switching Costs.

\bibitem{metcalfe_law_1995}
Bob Metcalfe:
\newblock Metcalfe's law: A network becomes more valuable as it reaches more
  users.
\newblock {\em InfoWorld\/}.

\bibitem{davis_power_2007}
Cornelia Davis:
\newblock The power of participation (Collaboration 2.0, Reed's Law and
  Metcalfe's Law)
\newblock
  \url{http://corneliadavis.com/blog/2007/02/09/the-power-of-participation-collaboration-20-reeds-law-and-metcalfes-law/}.

\bibitem{briscoe_metcalfes_2006}
Bob Briscoe, Andrew Odlyzko, and Benjamin Tilly:
\newblock Metcalfe's Law is Wrong - {IEEE} Spectrum
\newblock
  \url{https://spectrum.ieee.org/computing/networks/metcalfes-law-is-wrong}.

\bibitem{currier_network_2018}
James Currier:
\newblock The Network Effects Manual: 13 Different Network Effects (and
  counting)
\newblock
  \url{https://medium.com/@nfx/the-network-effects-manual-13-different-network-effects-and-counting-a3e07b23017d}.

\bibitem{hagiu_strategic_2015}
Andrei Hagiu:
\newblock Strategic Decisions for Multisided Platforms.
\newblock {\em MIT Sloan Management Review\/}.

\bibitem{hagiu_information_2014}
Andrei Hagiu and Hanna Halaburda:
\newblock Information and two-sided platform profits.
\newblock {\em International Journal of Industrial Organization\/}.

\bibitem{schmalensee_platform_2014}
Richard Schmalensee:
\newblock An Instant Classic: Rochet \& Tirole, Platform Competition in
  Two-Sided Markets.
\newblock {\em Competition Policy International\/}.

\bibitem{walter_4_major_2016}
Matthias Walker:
\newblock 4 Major Platform Business Models
\newblock
  \url{https://medium.com/platform-innovation-kit/4-major-platform-business-models-447e185d5ab5}.

\bibitem{schreieck_governing_2011}
Maximilian Schreieck, Christoph Hakes, Manuel Wiesche, and Helmut Krcmar:
\newblock Governing Platforms in the Internet of Things.
\newblock {\em Springer-Verlag\/}.

\bibitem{evans_platform_2011}
Daniel Evans:
\newblock {\em Platform Economics: Essays on Multi-Sided Businesses}
\newblock Createspace Independent Publishing Platform.

\bibitem{taeuscher_understanding_2017}
Karl T\"{a}uscher and Sven~M. Laudien:
\newblock Understanding platform business models: A mixed methods study of
  marketplaces.
\newblock {\em European Management Journal\/}.

\bibitem{parker_platform_2016}
Geoffrey~G. Parker, Marshall~W. Van~Alstyne, and Sangeet~Paul Choudary:
\newblock {\em Platform Revolution}
\newblock W. W. Norton \& Company, Inc.

\bibitem{kollmann_virtual_1999}
Tobias Kollmann:
\newblock Virtual Marketplaces: Building Management Information Systems for
  Internet Brokerage.
\newblock {\em Virtual Reality\/}.

\bibitem{hagiu_multi_platforms_2006}
Andrei Hagiu:
\newblock Multi-Sided Platforms: From Microfoundations to Design and Expansion
  Strategies.
\newblock {\em Harvard Business School\/}.

\bibitem{schmueck_democratization_2019}
Kilian Schm\"{u}ck:
\newblock The Democratization of the Platform Economy
\newblock
  \url{https://medium.com/share-charge/the-democratization-of-the-platform-economy-c4a5907bc7cc/}.

\bibitem{herd_towards_2018}
Benjamin Herd, Nik Scharmann, and Steve Phelps:
\newblock Towards the Model-Based Analysis and Design of Decentralised
  Economies of Things.
\newblock {\em King's Research Portal\/}.

\bibitem{choudary_dangers_2017}
Sangeet~Paul Choudary:
\newblock The Dangers of Platform Monopolies Knowledge
\newblock
  \url{https://knowledge.insead.edu/blog/insead-blog/the-dangers-of-platform-monopolies-6031}.

\bibitem{einav_peer_markets_2015}
Liran Einav, Chiara Farronato, and Jonathan Levin:
\newblock Peer-to-Peer Markets.
\newblock {\em NBER Working Paper No. 21496\/}.

\bibitem{baligh_vertical_market_1964}
H.~H. Baligh and L.~E Richartz:
\newblock An analysis of vertical market structures.
\newblock {\em Management Science\/}.

\bibitem{kollmann_elektronische_1998}
Tobias Kollmann:
\newblock Elektronische Marktpl\"{a}tze: Spielregeln f\"{u}r Betreiber
  virtueller Handelsr\"{a}ume.

\bibitem{kollmann_matching_2005}
Tobias Kollmann:
\newblock The matching function for electronic market places: determining the
  probability of coordinating of supply and demand.
\newblock {\em Int. J. Electronic Business\/}.

\bibitem{kollmann_toward_2019}
Tobias Kollmann, Simon Hensellek, Katharina de~Cruppe, and Andr\'{e} Sirges:
\newblock Toward a renaissance of cooperatives fostered by Blockchain on
  electronic marketplaces: a theory-driven case study approach.
\newblock {\em Topical Collection on Potential and Limits of Blockchain
  Technology for Networked Businesses\/}.

\bibitem{kollmann_e-business_2019}
Tobias Kollmann:
\newblock {\em E-Busines}
\newblock Gabler Verlag.

\bibitem{brandom_monopoly_2018}
Russell Brandom:
\newblock The monopoly-busting case against Google, Amazon, Uber, and Facebook
\newblock
  \url{https://www.theverge.com/2018/9/5/17805162/monopoly-antitrust-regulation-google-amazon-uber-facebook}.

\bibitem{akerlof_lemons_1970}
George~A. Akerlof:
\newblock The Market for "Lemons": Quality Uncertainty and the Market
  Mechanism.
\newblock {\em The Quarterly Journal of Economics\/}.

\bibitem{walker_how_2017}
Tim Walker:
\newblock How much ...? The rise of dynamic and personalised pricing
\newblock
  \url{https://www.theguardian.com/global/2017/nov/20/dynamic-personalised-pricing}.

\bibitem{gonzaga_personal_pricing_2018}
Piedro Gonzaga:
\newblock Personalised Pricing in the Digital Era.
\newblock {\em OECD Competition Division\/}.

\bibitem{duran_trivago_2020}
Paulina Duran:
\newblock Trivago misled customers by hiding best deals: Australian court
\newblock
  \url{https://www.reuters.com/article/us-trivago-australia-court/trivago-misled-customers-by-hiding-best-deals-australian-court-idUSKBN1ZK0MG}.

\bibitem{hempel_yelp_2020}
Klaus Hempel:
\newblock Yelp darf Bewertungen aussortieren
\newblock \url{https://www.tagesschau.de/bgh-yelp-urteil-101.html}.

\bibitem{srinivasan_decentralization_2017}
Balaji~S Srinivasan and Leland Lee:
\newblock Quantifying Decentralization
\newblock
  \url{https://news.earn.com/quantifying-decentralization-e39db233c28e}.

\bibitem{buterin_decentralization_2017}
Vitalik Buterin:
\newblock The Meaning of Decentralization
\newblock
  \url{https://medium.com/@VitalikButerin/the-meaning-of-decentralization-a0c92b76a274}.

\end{thebibliography}
\selectlanguage{english}%

\end{document}